\documentclass{iopart}

\usepackage{iopams}
\usepackage{epsfig}
\usepackage{graphics}

\begin{document}

\title{Polyelectrolyte Bundles}

\author{H. J. Limbach, M. Sayar, and C. Holm\footnote[1]{To whom
    correspondence should be addressed}}

\address{Max-Planck-Institut f\"ur Polymerforschung, Ackermannweg 10, 55128
  Mainz, Germany}

\begin{abstract}
Using extensive Molecular Dynamics simulations we study the
behavior of polyelectrolytes with hydrophobic side chains, which are
known to form cylindrical micelles in aqueous solution. We
investigate the stability of such bundles with respect to
hydrophobicity, the strength of the electrostatic interaction, and
the bundle size. We show that for the parameter range relevant for
sulfonated poly-para-phenylenes (PPP) one finds a stable finite
bundle size. In a more generic model we also show the influence of
the length of the precursor oligomer on the stability of the
bundles. We also point out that our model has close similarities
to DNA solutions with added condensing agents, hinting to the
possibility that the size of DNA aggregates is under certain
circumstances thermodynamically limited.
\end{abstract}

\eads{\mailto{\{limbach, sayar, holm\}@mpip-mainz.mpg.de}}

\section{Introduction} %
Aggregates of charged semiflexible polymers have been experimentally observed
in a variety of systems, where the most well known examples are probably
toroidally shaped bundles in DNA solutions with added condensing agents, such
as multivalent counterions \cite{bloomfield91a,bloomfield96a,conwell03a}.
Another polyelectrolyte system of synthetic nature, consisting of
poly(para-phenylene) (PPP) oligomers
\cite{bockstaller00a,bockstaller01a,wegner03a}, which are short rodlike
charged objects, forms cylindrical micelles due to its hydrophobic side
chains. This model can be well controlled chemically, and can be used as a
synthetic model system to study nano-aggregation of charged filaments.  An
additional advantage of PPPs is that they allow one to tune experimentally the
electrostatic and hydrophobic interactions separately. Although the physical
origin of both aggregation mechanisms is different, in a first approximation
they can be regarded as being the result of a short range attraction. This
interaction is, in the case of DNA, the result of short range ionic
correlations of the multivalent counterions \cite{rouzina96a}, whereas in the
case of the PPPs it is due to short range interactions of the hydrophobic side
chains. The aggregate size in both cases is determined by the competition
between the surface tension (due to short range attractions), the repulsive
self-energy of the backbone charges, and the entropic degrees of freedom of
the counterions. For DNA it has been argued
\cite{ha98a,levin02a,ha99c,stilck02a} that the observed finite size of the DNA
aggregates is due to kinetic problems, i.e. electrostatic barrier formation,
and not set by thermodynamic properties of the system. Experiments performed
with viral DNA support this conjecture at least for the investigated systems
\cite{lambert00a}. However, recently Henle and Pincus \cite{henle03_priv} have
argued, using a charged rod model interacting via a short range attraction,
that, depending on the actual parameters of the system, either finite or
infinite bundle sizes should be possible. Treating the systems as consisting
of sticky charged rods brings up the analogy to the Rayleigh split of a
charged hydrophobic droplet. This problem has been solved in a mean-field
model already by Deserno \cite{deserno01f}.  He observes that the droplet size
is always finite, if the counterions cannot penetrate into the droplet, but
that allowing counterion penetration leads either to finite or infinite
droplet sizes, depending on the parameters. There have been relatively few
simulations on bundle formations, notable exception being the simulations by
Stevens \cite{stevens99a,stevens01a}, showing the possibility that multivalent
ions alone can lead to bundle formations, and the work of Borukhov et al.
\cite{borukhov02a}, which investigates two-rod systems bundled via short range
mobile linker interactions. In the present work we investigate the bundle
formation for a system of charged semi-flexible polymers with short range
interactions for two models, and demonstrate clearly the existence of finite
size aggregates in both cases.

\section{Simulation method}
\subsection{PPP \label{ppp-method}} %
\begin{figure}[hb]
\begin{centering}
  \includegraphics[width=0.9\linewidth]{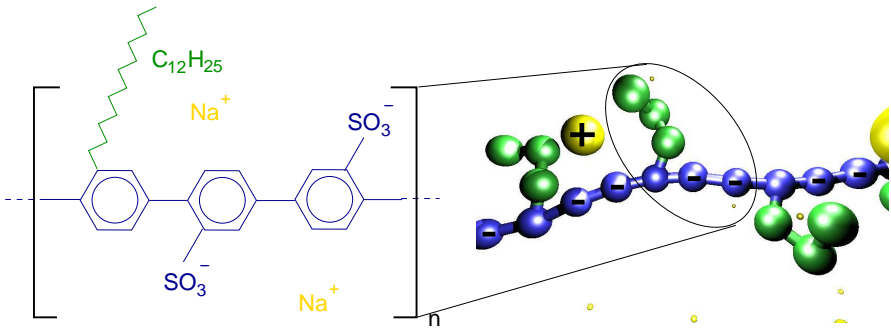}
  \caption{Mapping of a PPP monomer onto the bead spring model.\label{coarse_graining} }
\end{centering}
\end{figure}

The PPP oligomer is described with a bead spring model, whose mapping is shown
in figure.~\ref{coarse_graining}.  Each phenyl ring along the semi-flexible
backbone is represented by a spherical bead. Two of them carry one negative
unit charge $e_o$.
 The chain length is $N_m=61$ beads representing 20 PPP
monomers. The hydrophobic side chain is modeled as a flexible
chain containing 3 beads. The counterions are simulated explicitly
as charged spheres carrying one unit charge. The solvent is taken
into account implicitly represented as a background with
dielectric constant $\epsilon$. The simulations are performed in an
NVT ensemble. The temperature is maintained via a Langevin
thermostat.

All particles interact via a repulsive Lennard-Jones potential with
$r_{cut}=2^{1/6}\sigma$ and $\epsilon_{LJ}=1 k_B T$
(eq.~\ref{lennard-jones}) and charged particles interact via an
unscreened Coulomb potential (eq.~\ref{coulomb}). Bonds are realized
with a FENE~\cite{grest86a} potential with $k_F=7 k_B T$ and a cutoff
$R_F = 2 \sigma$ (eq.~\ref{fene}). The PPP backbone is semi-flexible
which is modeled with a bond angle potential with $k_{\theta}=30 k_B
T$ (eq.~\ref{angle}). The influence of the solvent on the hydrophobic
side chains is taken care of with an attractive Lennard-Jones
potential between these monomers (eq.~\ref{lennard-jones}). Here the
cutoff is $r_{cut}=2.5\sigma$.

\begin{equation}
\label{lennard-jones} U_{LJ}(r)= 4 \epsilon_{LJ} \left(
\left(\frac {\sigma} {r} \right)^{12} - \left(\frac {\sigma} {r}
\right)^{6}\right) \quad  $for$ \quad r < r_{cut}$$
\end{equation}

\begin{equation}
\label{coulomb} U_{C}(r_{ij})= \ell_B T \frac {q_i q_j} {r_{ij}}
\end{equation}

\begin{equation}
\label{fene} U_{F}(r)= \frac {1} {2} k_{F} R^2_{F}
\ln\left(1-\left(\frac {r} {R_{F}}\right)^2\right) \quad $for$ \quad r < R_F$$
\end{equation}

\begin{equation}
\label{angle} U_{\theta}(r)= k_{\theta} \left( 1 -\cos \theta
\right)
\end{equation}

The parameters $\epsilon_{LJ}$ for the attractive Lennard-Jones
interaction and the Bjerrum length $\ell_B = e_0^2/(4 \pi \epsilon k_B
T)$ have been varied. Due to the large kinetic barriers involved in
the formation of a polyelectrolyte bundle, we study only its possible
break up by starting the simulations with a preformed bundle in the
middle of a spherical simulation cell. The cell radius is given by the
particle density which is $\rho=1.0 \times 10^{-4} \sigma^{-3}$. The
equilibration is done in a two step process.  First we constrain the
backbone monomers of the PPPs in space and let the hydrophobic hairs
and counterions equilibrate for 100 000 MD steps. Then we release the
backbone monomers and let the total system equilibrate. The bundle is
simulated for 2 000 000 MD steps. If the bundle is stable over the
whole simulation time it is called stable in this paper. We are well
aware that this procedure only proves that the observed bundles are at
least in a metastable state and cannot prove that we are really in the
global minimum of the free energy. However, we only want to
demonstrate the existence of {\it finite} aggregate sizes, and
therefore our method suffices, see e.g.  Ref. \cite{borukhov02a}
where the same simulation technique has been applied to the case of a
two bundle system.

\subsection{Generic bundles} %
Further simplification of this model can be achieved by taking the zero size
limit of the hydrophobic side chains and making instead all backbone monomers
hydrophobic, which is exactly the model Henle and Pincus suggested
\cite{henle03_priv}. Each bead carries a positive unit charge. The beads
interact with all other charges via an unscreened Coulomb
interaction (eq.~\ref{coulomb}). The beads are connected by a FENE potential
(eq.~\ref{fene}) and furthermore the oligomer is semi-flexible, where the
stiffness of the chain can be tuned by a bond angle potential
(eq.~\ref{angle}). The hydrophobic interactions due to side chains are also
represented via short range interactions of these beads, where the
Lennard-Jones+Cosine potential (eq.~\ref{cosine}) is used~\cite{soddeman01a}.
This potential enables the use of large $\epsilon_{LJ}$ values, while the
function and it's first derivative are continuous at the cut off radius. The
cut off radius is chosen to be extremely small , $r_{cut}=1.5 \sigma $, so
that the counterions can not penetrate into the bundle without breaking the
short range interaction among the polyelectrolyte beads. All other interaction
parameters are as described in section~\ref{ppp-method}.

\begin{equation}
\label{cosine} U_{COS}(r)=\frac {\epsilon_{LJ}} {2}  \left(\cos
\left(\alpha r^2 + \beta\right) -1\right) \quad $for$ \quad r_{min} \le r <
r_{cut}$$
\end{equation}

\section{Results and discussions} %
\subsection{PPP}

\begin{figure*} [ht]
    \begin{centering}
        \includegraphics[scale=0.9]{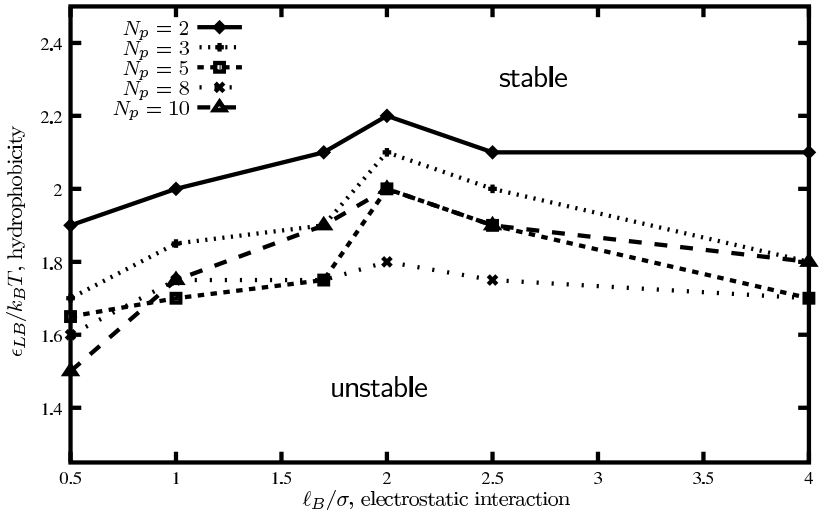}
        \caption{\label{phase_space} $\epsilon_{LJ}$-$\ell_B$ phase
          diagram for bundles with $N_p$=2, 3, 5 ,8 ,10. }
    \end{centering}
\end{figure*}

For the PPP model we are interested in the influence of the
hydrophobicity and the strength of the electrostatic interaction
on the stability of the PPP micelles. We simulated PPP bundles of
different aggregate sizes and varied systematically the parameters
for the hydrophobicity $\epsilon_{LJ}$ and the strength of the
electrostatic interaction $\ell_B$. The simulated bundle sizes $N_p$
are 2, 3, 5, 8 and 10. In Figure~\ref{phase_space} we show the
stable and unstable regions in a phase diagram.

The phase boundary for bundles with different sizes are shown as
lines connecting simulations where the bundles are marginally
stable. Above the line bundles with that specific size are stable
(actually, only at least meta-stable), below unstable.

Looking at the phase boundary for a given bundle size, e.g. $N_p=5$, with
increasing $\ell_B$ the stability of the bundle first decreases. This means
one needs a larger hydrophobicity to get a stable bundle. This can be
explained by the increasing electrostatic repulsion of the charged backbones.
At $\ell_B/\sigma \simeq 2$ the stability curve shows a maximum. Further increase of
$\ell_B$ leads to more stable bundles. This non-monotonic form resembles the
behavior of the extension of flexible polyelectrolyte chains both in good
\cite{stevens95a} and poor solvent
\cite{micka00a,limbach02a,limbach02c,limbach03a}. The increasing stability at
high $\ell_B$ values is due to the increased density (condensation) of
counterions close to the bundle which renormalize the effective charge and can
also mediate attractive interactions at sufficiently large values of $\ell_B$.

In Figure~\ref{phase_space} we can also compare the stability of bundles with
different size. For $\ell_B=0.5\sigma$ where the electrostatic interaction
does not play a significant role, we see that for the investigated regime the
range of $\epsilon$-values yielding stable bundles increases with increasing
bundle size. This behavior changes drastically when one looks at $\ell_B$
values larger than $1.0\sigma$.  Here we find at $\epsilon$-values between 1.7
and 1.9 that bundles which contain between 5 and 8 PPPs are more stable than
larger or smaller $N_p$ values. This implies also that one can find situations
with a finite stable bundle size. Note that we observe this behavior in the
relevant experimental region, namely $\ell_B=1.7\sigma$ for PPP. We can
explain this finding by an electrostatic instability in analogy to the
Rayleigh instability of charged droplets \cite{rayleigh1882a,deserno01f},
which has been shown to be valid also for linear charged polymers
\cite{kantor94a,dobrynin96a}. When one further increases $\ell_B$, the marginal
stable bundle size decreases more and more, and the stability lines for all
bundles containing more than 3 PPPs merge.  We expect that for even larger
$\ell_B$ values one will find an infinite bundle size.

\begin{figure*} [ht]
    \begin{centering}
        \includegraphics[width=\linewidth]{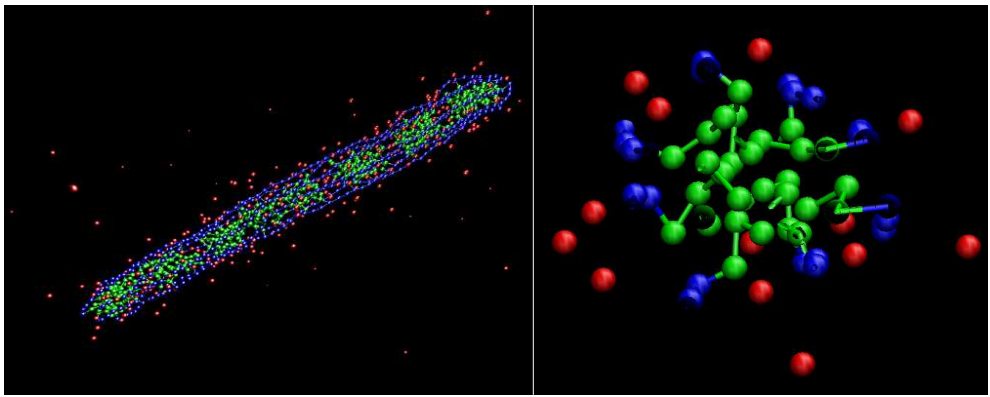}
        \caption{\label{bundle-snap} The snapshot of a polyelectrolyte
        bundle (left) together with a cross section (right) is shown for
        $N_p=8$, $\epsilon_{LJ}=1.7 k_B T$ and $\ell_B=1.7\sigma$  }
    \end{centering}
\end{figure*}

Finally we investigated the counterion distribution inside and in
the close vicinity of a bundle. In Figure~\ref{bundle-snap} we
show a snapshot of a polyelectrolyte bundle together with its
cross section. Note that the counterions are able to penetrate
also inside the bundle. This is important for the theoretical
approach to the Rayleigh instability of charged droplets as it is
introduced by Deserno~\cite{deserno01f}.

\subsection{Generic bundles}
Also for the generic bundle model as discussed above we find the
formation of finite size aggregates. In order to study the
contribution of various effects (e.g. strength of Coulomb and
Lennard-Jones interactions, stiffness of the chains) the
stability of a bundle of certain size can be tested as follows.
A cylindrical bundle is preformed by aligning the backbone of
the polymers along the axis of the cylinder. The cross-section of
the bundle is chosen as a close-packed structure on a triangular
lattice to minimize the surface area. Unlike the PPP polymers in
the previous section, where the hydrophobic side chains where
explicit, in this more generic model the hydrophobic interaction
is also associated with the backbone beads. Therefore the polymers
are not packed like a classical micelle. The lattice spacing for
this initial configuration is chosen as the minimum of the
Lennard-Jones + Cosine potential. The counterions are placed in a
cylinder enclosing the bundle, where the radius of this cylinder
is large enough to provide sufficient freedom during the
relaxation of the counterions. The equilibration is started with
the counterions, which are set free to move within this cylinder
surrounding the bundle. However, it is important to note that the
counterions cannot penetrate into the bundle at this stage due to
the close packing of the polymers. After the counterions are
equilibrated, the polymers are set free, and a brief equilibration
run is performed where the polymers are also confined within a
cylinder tightly enclosing the polymers, so that the bundle does
not fall apart at this stage. Next the cylinder constraints are
removed, and the system is integrated for $2.0 \times 10^6$ time
steps. During the simulation, depending on the relative value of
the interaction parameters, the bundle either remains as a single
aggregate or splits up into smaller aggregates.

\begin{figure*} [ht]
    \begin{centering}
        \includegraphics[scale=0.9]{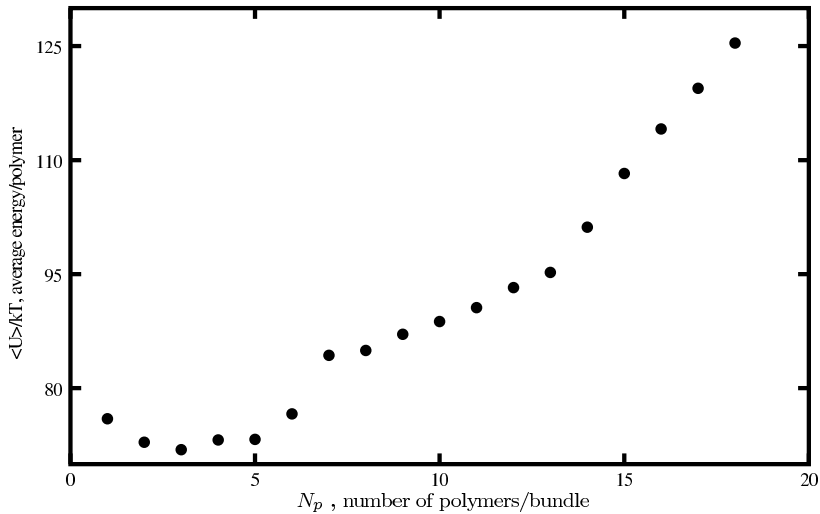}
        \caption{\label{lj4_aveU} The average internal energy per
        polymer as a function of number of polymers ($N_p$) in the
        startup bundle.}
    \end{centering}
\end{figure*}

In Figure \ref{lj4_aveU} the biggest stable bundle size obtained
from the simulation is shown as a function of the number of
polymers in the initial bundle. The polymers are $N_m$=10 beads
long, and $\epsilon_{LJ} =4.0 k_BT$ and $l_B  =2.0 \sigma$. The
stiffness of the chains is kept fixed ($k_\theta  = 10 k_BT$). Up
to six polymers per startup bundle, the structure does not
disintegrate and the bundle remains as a single entity. However,
for initial aggregates with more than $N_p$=6 polymers, the
aggregate splits into two or more bundles. In Figure
\ref{lj4_aveU} the biggest size among these bundles is plotted.
The stable equilibrium bundle size for this set of parameters is
$N_p \approx5$. However, due to the finite system size the equilibrium
aggregate size obtained via this method must be taken cautiously.
\begin{figure*} [ht]
    \begin{centering}
        \includegraphics[scale=0.9]{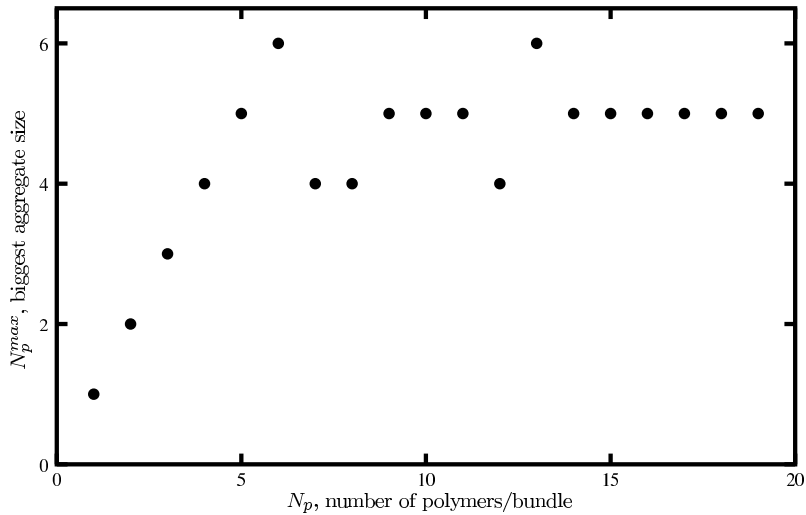}
        \caption{\label{lj4_aveN} The biggest stable aggregate size
        ($N_p^{max}$) as a function of number of polymers ($N_p$)
        in the startup bundle.}
    \end{centering}
\end{figure*}

Another approach to obtain the average bundle size is to look at
the internal energy of a bundle with counterions. The bundle is
formed as in the previous case; however, in this simulation the
bundle is constraint to remain as a single aggregate. In other
words, the polymers are fixed throughout the simulation, and only
the counterions are free to move. The average internal energy of
the system as a function of the number of polymers per aggregate
is plotted in Figure \ref {lj4_aveU} for the same interaction
parameters as in the previous case. In these constrained bundle
simulations a minimum in the internal energy of the system is
observed for $N_p$=3. Initially as the number of polymers per
bundle increases, the internal energy of the system decreases as a
result of decreasing surface energy. However, above the average
aggregate size the electrostatic energy of the system increases
dramatically. The energy distribution for bundles smaller than
$N_p$=7 polymers is relatively flat. The sudden increase in the
internal energy for the $N_p$=7 polymer aggregate is due to the
complete isolation of the central polymer from the surrounding
counterions. This is an artifact of fixing the polymers during the
simulation. It is important to note that only the internal energy
of the system is taken into account in this analysis. However, the
entropy of the counterions could also play a dominant role in
determining the average aggregate size for these polyelectrolytes
with hydrophobic interactions. The slight difference in the
equilibrium aggregate size obtained with these two methods
presented above can be explained by the entropic contribution
which is neglected in the fixed bundle simulations.

\begin{figure*} [ht]
    \begin{centering}
        \includegraphics[scale=.3]{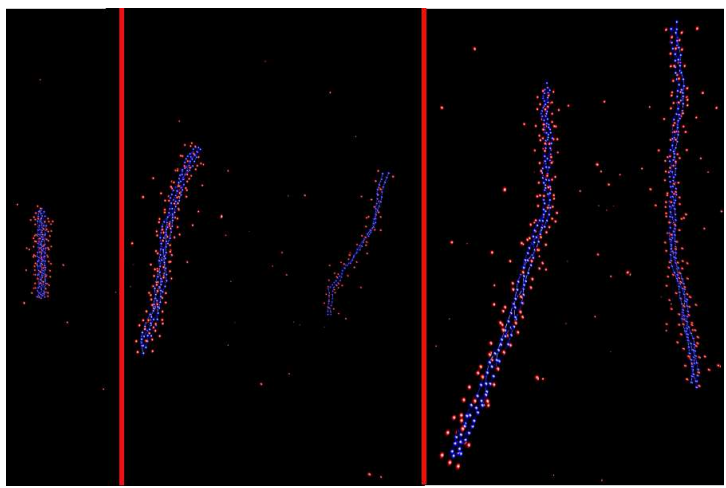}
        \caption{\label{length} Snapshots from simulations of a 6
        polymer bundle with polymer lengths $N_m$=20, $N_m$=40,
        and $N_m$=60 from left to right, respectively. }
      \end{centering}
\end{figure*}

The length and charge density of the precursor polymers effect the fraction of
condensed ions. Even though Poisson-Boltzmann theory yields a good
approximation for the case of infinitely long polymers \cite{deserno00a}, for
finite length rods there are no straight-forward solutions. In order to gain
further inside into the finite length effects, we have performed bundle
simulations for lengths ranging from $N_m$=10 to $N_m$=60 monomers. Snapshots
from the equilibrated systems of $N_m$=20, $N_m$=40, and $N_m$=60 are shown in
Figure \ref{length}. For bundles with less than $N_p$=6 polymers all simulated
polymer lengths yield stable aggregates, no splitting is observed. On the
other hand, for bundles of 6 polymers even though the polymers of length
$N_m$=10 and $N_m$=20 monomer length provide stable aggregates, for longer
polymer lengths the bundle splits up.

\begin{figure*} [ht]
    \begin{centering}
        \includegraphics[scale=0.9]{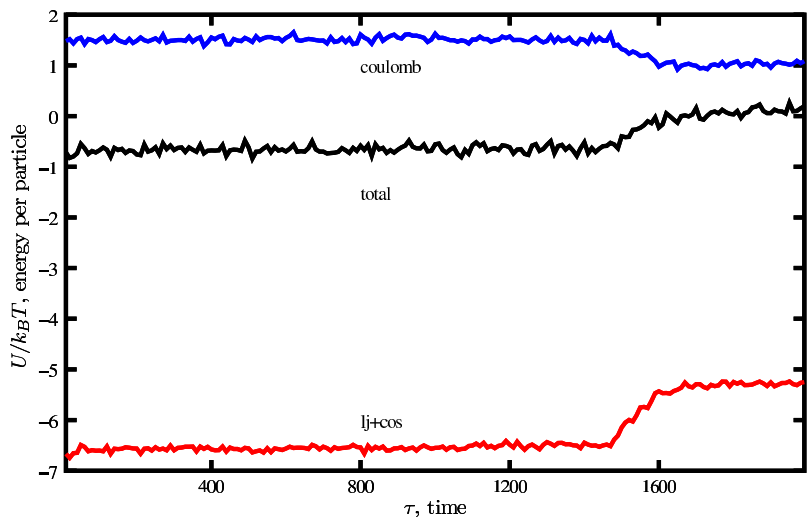}
        \caption{\label{split-energy} The total internal energy (middle),
        Coulomb (top) and Lennard-Jones (bottom) contributions per monomer
        for bundles of 6 polymers during the simulation.}
    \end{centering}
\end{figure*}

The internal energy and the contributions from Coulombic and
Lennard-Jones potentials per particle for bundles of $N_p$=6
polymers with polymer length of $N_m$=60 monomers is shown in
Figure \ref{split-energy}. Upon splitting, the internal energy of
the bundle increases by almost one $k_BT$ per particle. The split
takes place rather abruptly in a late stage of the simulation,
which suggests the presence of a high energy barrier for the
split. The split of the bundle into two small bundles is
unfavorable in terms of the short range Lennard-Jones interactions
since the surface energy increases. On the other hand for the
Coulombic interactions the answer is more complicated. The
electrostatic self energy of the bundle decreases due to split,
since the like-charged polymers repel each other. However, since
the split-bundles attract less counterions the favorable
electrostatic interactions among the polymers and counterions is
also lost. For $N_p$=6 polymer bundle upon splitting the Coulombic
contribution to the internal energy decreases.

\begin{figure*} [ht]
    \begin{centering}
        \includegraphics[scale=0.9]{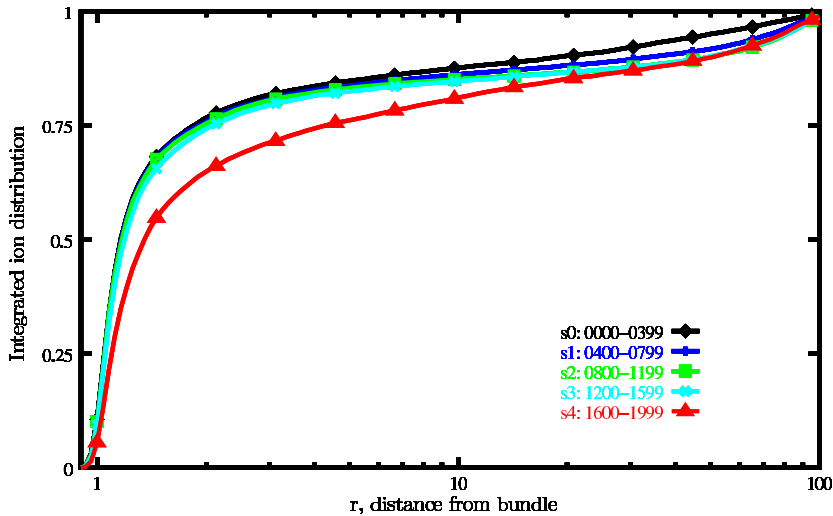}
        \caption{\label{split-ion} Integrated counterion distribution
            as a function of radial distance from the 6 polymer bundle.
            The distributions are calculated for five different time
            periods during the simulation. Time periods from top curve to the bottom one: s0 (0-399) - s4 (1600-1999). Time period s4 corresponds to the distribution after the bundle splits up. }
    \end{centering}
\end{figure*}

The increase in the internal energy of the system can be advantageous for the
system only if the release of the counterions sufficiently increases the
entropy to compensate for the increase in internal energy. In Figure
\ref{split-ion} the integrated counterion distribution for the $N_p$=6 polymer
bundle is given for five time intervals during the simulation. For each
counterion the distance to the closest monomer is chosen for the radial
distribution function. Prior to the split (s0-s3) the counterion distribution
does not change. Upon splitting up into two bundles the fraction of ions close
to the bundles decreases dramatically. Therefore we can conclude that the
splitting of these long polymer bundles are driven by the entropy of the
released counterions.

\section{Conclusions} 
We have shown that charged semi-flexible polyelectrolytes interacting
via two kinds of hydrophobic short range interactions have parameter regions
where finite aggregate sizes exist. For large values of $l_B$ the bundle size
increases, and in the limit of large $l_B$ we observe the expected trend to
form infinite bundles due to counterion crystallization. Furthermore, we could
demonstrate that the release of counterions during a bundle split can
considerably lower the free energy of the total system.

The stability curve for different bundle sizes as a function of $l_B$ shows
similarities to the Rayleigh instability, i.e., for an increase in $l_B$ one
needs an increased $\epsilon_{LJ}$. Another observation is that the stability
line varies non-monotonically with Bjerrum length, which resembles the
non-monotonic extension behavior linear polyelectrolytes show if the Bjerrum
length is increased. We relate this to the fact that the electrostatic
self-energy of the bundle does not increase with $l_B$ after $l_B \approx
2\sigma$ due to counterion-backbone ion correlations. The degree of
polymerization shows also an effect on bundle size, which has been observed in
experiments \cite{bockstaller00a,bockstaller01a} on PPPs as well; however, our
limited data does not allow us to draw definite conclusions about the
underlying mechanism.

One important point to note is the similarities of our model with the observed
trends in DNA condensation experiments which also find finite aggregate sizes.
The applicability of our model to those experiments rests basically on two
assumptions. First, we assume that the strength of the short range attraction
does not change with external parameters, and second, presently we have not
gauged the interaction strength to that originating from multivalent
counterion interactions. This important piece of work is left for future
investigations.  However, it appears plausible that at least for some
parameter regions of biological charged polymers both assumptions are
justified, so that our results can be used to provide a mechanism for finite
bundle sizes.

Our results also demonstrate that the mean-field theory of Deserno
\cite{deserno01f} for a charged hydrophobic droplet looks qualitatively
correct. Extending the Deserno analysis to charged sticky rods does not change
the qualitative behavior \cite{tamashiro03_pre}.
%
\section{Acknowledgments}
We would like to thank M. Deserno, P. Pincus, H. Schiessel, and M.
Tamashiro for stimulating discussions.  This work was supported by
the ``Zentrum f\"ur Multifunktionelle Werkstoffe und
Miniaturisierte Funktionseinheiten", grant BMBF 03N 6500, and the
DFG within the SFB 625 and grant HO-1108/11-1.
\section*{References}

\bibliographystyle{prsty}
\bibliography{../bibtex/polyelectrolyte}
\end{document}